\documentstyle[aps]{revtex}
\input{psfig} 
\begin{document}
%\draft
\vspace{2.in}
\title{ Kaon electroweak form factors in the light-front 
quark model} 
\author{ Ho-Meoyng Choi and Chueng-Ryong Ji}
\address{
Department of Physics,
North Carolina State University,
Raleigh, N.C. 27695-8202}
%\date{\today}
\maketitle
%\narrowtext
\begin{abstract}
We investigate the form factors and decay rates for the semileptonic decays 
of the kaon($K_{l3})$ using the light-front quark model.   
The form factors $f_{\pm}(q^{2})$ are calculated in $q^{+}=0$ frame 
and analytically continued to the time-like region, $q^{2}>0$.
Our numerical results for the physical observables, 
$f_{-}/f_{+}|_{q^{2}=m^{2}_{l}}=-0.38$, 
$\lambda_{+}=0.025$(the slope of $f_{+}$ at   
$q^{2}=m^{2}_{l}$), $\Gamma(K^{0}_{e3})=(7.30\pm 0.12)\times10^{6}s^{-1}$,
and $\Gamma(K^{0}_{\mu3})=(4.57\pm0.07)\times10^{6}s^{-1}$ are quite  
comparable with the experimental data and other theoretical model 
calculations. The non-valence contributions from $q^{+}\neq 0$ frame are 
also estimated.
\end{abstract}
\newpage
\baselineskip=20pt
\section{Introduction }
Even though there have been a lot of analyses on the heavy-to-heavy and
heavy-to-light form factors for weak transitions from a pseudoscalar
meson to another pseudoscalar meson  within the light-front quark
model(LFQM)\cite{OD1,OD2,Jaus,Demchuk,Grach,Simula,Cheng,Melikhov},
the light-to-light weak form factor analysis such as $K_{l3}$ has not
yet been studied in LFQM. However,  
the analysis of semileptonic $K_{l3}$ decays comparing with the 
experiment\cite{data} has been provided by many other theoretical  
models, $e.g.$, the chiral perturbation theory(CPT)\cite{Roos,Gasser},   
the effective chiral Lagrangian approach\cite{CL}, the vector meson  
dominance\cite{VMD}, the extended Nambu-Jona-Lasino model\cite{Andrei},
Dyson-Schwinger approach\cite{Yuri} and other quark model\cite{ISGW2,Isgur}.   
Thus, in this work, we use LFQM to analyze both form factors of the 
$K_{l3}$ decays, $i.e.$, $f_{+}$ and $f_{-}$, and compare with the 
experimental data as well as other theoretical models. 

In the LFQM calculations presented in Refs.\cite{Demchuk,Grach,Simula,Cheng}, 
$q^{+}\neq 0$ frame has been used to calculate the weak decays in 
the time-like region $m^{2}_{l}\leq q^{2}\leq (M_{i}-M_{f})^{2}$, with 
$M_{i[f]}$ and $m_{l}$ being the initial[final] meson mass and the 
lepton($l$) mass, respectively.
However, when the $q^{+}\neq 0$ frame is used, the inclusion of the
non-valence contributions arising from quark-antiquark pair
creation(``Z-graph") is inevitable and this inclusion may be 
very important for heavy-to-light and light-to-light decays. 
Nevertheless, the previous
analyses\cite{Demchuk,Grach,Simula,Cheng} in $q^{+}\neq 0$ frame
considered only valence contributions neglecting non-valence 
contributions. In this work, we circumvent this problem by calculating 
the processes in $q^{+}=0$ frame and analytically continuing to the
time-like region. The $q^{+}=0$ frame is useful because only 
valence contributions are needed. However, one needs to calculate  
the component of the current other than $J^{+}$ to obtain the form factor 
$f_{-}(q^{2})$. Since $J^{-}$ is not free from the zero-mode contributions 
even in $q^{+}=0$ frame\cite{Zm,Brodsky},
we use $J_{\perp}$ instead of $J^{-}$ to obtain $f_{-}$.
The previous works in Refs.\cite{OD1,OD2,Jaus} have considered only the  
$``+"-$component of the current which was not sufficient to obtain the form 
factor $f_{-}(q^{2})$. Furthermore, the light-to-light decays such as
$K_{l3}$ have not yet been analyzed, even though the calculation of $f_{-}$ 
for heavy-to-heavy and heavy-to-light decays has been made in 
Ref.\cite{Melikhov} using the dispersion formulations.
Thus, we analyze both currents of $J^{+}$ and $J_{\perp}$ for $K_{l3}$ 
decays using $q^{+}=0$ frame and analytically continue to the time-like region.
Our method of changing $q_{\perp}$ to $iq_{\perp}$ is not only simple to
use in practical calculations for the exclusive processes but also 
provides the identical results obtained by the dispersion formulations
presented in Ref.\cite{Melikhov}.

The calculation of the form factor $f_{-}(q^{2})$ is especially   
important for the complete analysis of $K_{l3}$ decays,
since the $f_{-}(q^{2})$ is prerequisite for the calculation of the physical
observables $\xi_{A}=f_{-}/f_{+}|_{q^{2}=m^{2}_{l}}$ and $\lambda_{-}$, the
slope of $f_{-}(q^{2})$ at $q^{2}=m^{2}_{l}$. 
We also estimate the non-valence contributions from $q^{+}\neq0$ frame by 
calculating only valence contributions from $q^{+}\neq0$ frame and comparing
them with those obtained from $q^{+}=0$ frame. Including the lepton mass 
effects for the $d\Gamma/dq^{2}$ spectrum of $K_{l3}$, we distinguish the 
decay rate of $K_{\mu3}$ from that of $K_{e3}$, where the contribution from 
$f_{-}$ is found to be appreciable for $\mu$ decays.     
 
Our model parameters summarized in Table I were
obtained from our previous analysis of quark potential model\cite{CJ1},
which provided a good agreement with the experimental data of
various electromagnetic properties of mesons such as
$f_{\pi},f_{K}$, charge radii of $\pi$ and $K$, and rates for radiative meson
decays $etc.$ As shown in Ref.\cite{CJ1}, the gaussian radial wave function 
$\phi(x,{\bf k}_{\perp})$ for our LF wave function 
$\Psi^{JJ_{z}}_{\lambda_{q},\lambda_{\bar{q}}}(x,{\bf k}_{\perp})
=\phi(x,{\bf k}_{\perp}){\cal R}^{JJ_{z}}_{\lambda_{q},\lambda_{\bar{q}}}
(x,{\bf k}_{\perp})$ is given by  
\begin{eqnarray}
\phi(x,{\bf k}_{\perp})&=& \sqrt{\frac{\partial k_{z}}{\partial x}}   
\biggl(\frac{1}{\pi^{3/2}\beta^{3}}\biggr)^{1/2}\exp(-k^{2}/2\beta^{2}),
\end{eqnarray}
where ${\partial k_{z}}/{\partial x}$ is the Jacobian of the variable
transformation $\{x,{\bf k}_{\perp}\}\rightarrow {\bf k}=
(k_{n},{\bf k}_{\perp})$. The spin-orbit wave function 
${\cal R}^{JJ_{z}}_{\lambda_{q},\lambda_{\bar{q}}}(x,{\bf k}_{\perp})$
is obtained by the interaction-independent Melosh transformation. 
The detailed description for the spin-orbit wave function can also 
be found in previous literatures\cite{OD1,OD2,Jaus,Grach,Simula,Cheng,CJ1}.

The paper is organized as follows. In Sec.II, we obtain the 
form factors of $K_{l3}$ decays in $q^{+}=0$ frame and analytically
continue to the time-like $q^{2}>0$
region by changing $q_{\perp}$ to $iq_{\perp}$ in the 
form factors. In Sec.III, our numerical results of the observables for 
$K_{l3}$ decays are presented and compared with the experimental data
as well as other theoretical results. 
Summary and discussion of our main results follow in Sec.IV.
In the Appendix A, we show the derivation of the matrix element of the 
weak vector current for $K_{l3}$ decays in the standard $q^{+}=0$ frame. 
In the Appendix B, the valence contribution in $q^{+}\neq 0$ frame is
formulated. 
\section{Weak Form factors in Drell-Yan frame}  
The matrix element of the hadronic current for $K_{l3}$  
can be parametrized in terms of two hadronic form factors as follows  
\begin{eqnarray}
<\pi|\bar{u}\gamma^{\mu}s|K>&=&
f_{+}(q^{2})(P_{K}+P_{\pi})^{\mu} + f_{-}(q^{2})(P_{K}-P_{\pi})^{\mu},
\nonumber\\
&=& f_{+}(q^{2})\biggl[(P_{K}+P_{\pi})^{\mu}
-\frac{M^{2}_{K}-M^{2}_{\pi}}{q^{2}}q^{\mu}\biggr]
+ f_{0}(q^{2})\frac{M^{2}_{K}-M^{2}_{\pi}}{q^{2}}q^{\mu},
\end{eqnarray}
where $q^{\mu}=(P_{K}-P_{\pi})^{\mu}$ is the four-momentum 
transfer to the leptons and $m^{2}_{l}\leq q^{2}\leq(M_{K}-M_{\pi})^{2}$. 
The form factors $f_{+}$ and $f_{0}$ are related to the exchange of $1^{-}$
and $0^{+}$, respectively, and satisfy the following relations:  
\begin{eqnarray}
f_{+}(0)&=& f_{0}(0),\hspace{.2cm}f_{0}(q^{2})= f_{+}(q^{2})
+ \frac{q^{2}}{M^{2}_{K}-M^{2}_{\pi}}f_{-}(q^{2}). 
\end{eqnarray}
Since the lepton mass is small except in the case of the $\tau$
lepton, one may safely neglect the lepton mass in the decay rate
calculation of the heavy-to-heavy and heavy-to-light transitions. 
However, for $K_{l3}$ decays, the muon($\mu$) mass is not negligible, 
even though electron mass can be neglected. Thus, including non-zero
lepton mass, the formula for the decay rate of $K_{l3}$ is given 
by\cite{Hwang}:
\begin{eqnarray} 
\frac{d\Gamma(K_{l3})}{dq^{2}}&=& \frac{G^{2}_{F}}{24\pi^{3}}|V_{us}|^{2}
K_{f}(q^{2})(1-\frac{m^{2}_{l}}{q^{2}})^{2}\nonumber\\
&\times&\biggl\{
[K_{f}(q^{2})]^{2}((1+\frac{m^{2}_{l}}{2q^{2}})|f_{+}(q^{2})|^{2} 
+ M^{2}_{K}(1-\frac{M^{2}_{\pi}}{M^{2}_{K}})^{2}\frac{3}{8}
\frac{m^{2}_{l}}{q^{2}}|f_{0}(q^{2})|^{2}\biggr\},
\end{eqnarray}  
where $G_{F}$ is the Fermi constant, $V_{q_{1}\bar{q_{2}}}$ is 
the element of the Cabbibo-Kobayashi-Maskawa mixing matrix and
the factor $K_{f}(q^{2})$ is given by 
\begin{eqnarray}
K_{f}(q^{2})&=& \frac{1}{2M_{K}}
\biggl[ (M_{K}^{2}+M_{\pi}^{2}-q^{2})^{2}-4M_{K}^{2}M_{\pi}^{2}
\biggr]^{1/2}.  
\end{eqnarray} 
Since our analysis will be performed in the isospin symmetry($m_{u}$=$m_{d}$)
but $SU_{f}(3)$ breaking($m_{s}\neq m_{u(d)}$) limit, we do not
discriminate between the charged and neutral kaon weak decays,
$i.e.$, $f^{K^{0}}_{\pm}=f^{K^{+}}_{\pm}$.
For $K_{l3}$ decays, the three form factor parameters,
$i.e.$, $\lambda_{+}, \lambda_{0}$ and $\xi_{A}$,
have been measured using the following linear parametrization\cite{data}:
\begin{eqnarray}
f_{\pm}(q^{2})&=& f_{\pm}(q^{2}=m^{2}_{l})\biggl(
1 + \lambda_{\pm}\frac{q^{2}}{M^{2}_{\pi^{+}}}\biggr),
\end{eqnarray}
where $\lambda_{\pm,0}$ is the slope of $f_{\pm,0}$ evaluated at
$q^{2}=m^{2}_{l}$ and $\xi_{A}=f_{-}/f_{+}|_{q^{2}=m^{2}_{l}}$. 
\begin{figure}
\psfig{figure=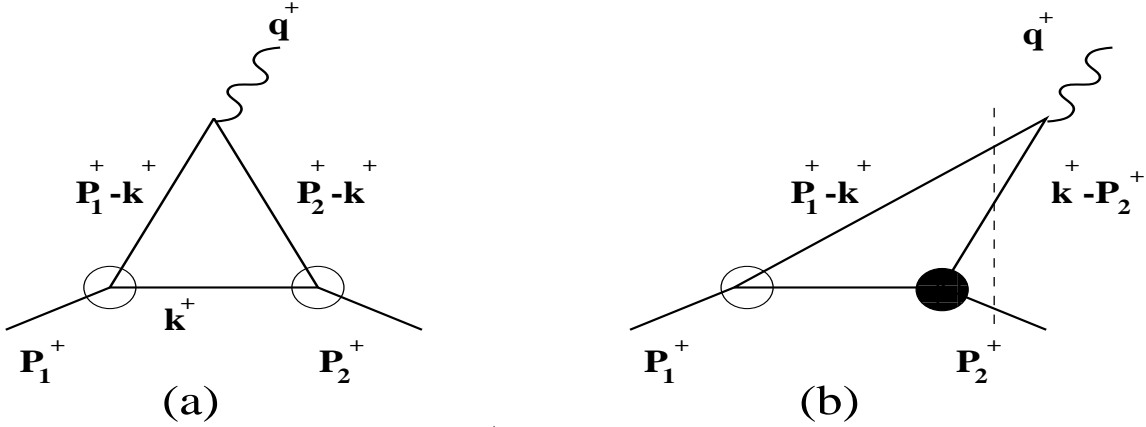,width=6in,height=2.2in}
\caption{The form factor calculation in $q^{+}\neq0$ frame requires both
the usual light-front triangle diagram(a) and the
non-valence(pair-creation) diagram(b). The vertical dashed line in (b) 
indicates the energy-denominator for the non-valence contribution.
While the white blob represents our LF wavefunction
$\Psi^{JJ_{z}}_{\lambda_{q},\lambda_{\bar{q}}}(x,{\bf k}_{\perp})$,
the modeling of black blob has not yet been made.}
\end{figure}
As shown in Fig.1, the quark momentum variables for $q_{1}\bar{q}\to 
q_{2}\bar{q}$ transitions in the standard $q^{+}=0$ frame are given by 
\begin{eqnarray}
p^{+}_{1}&=&(1-x)P^{+}_{1},\hspace{.5cm}
p^{+}_{\bar{q}}=xP^{+}_{1},\nonumber\\
{\bf p}_{1\perp}&=&(1-x){\bf P}_{1\perp} + {\bf k}_{\perp},\hspace{.5cm}
{\bf p}_{\bar{q}\perp}= x{\bf P}_{1\perp} - {\bf k}_{\perp},
\nonumber\\
p^{+}_{2}&=&(1-x)P^{+}_{2},\hspace{.5cm}
p'^{+}_{\bar{q}}=xP^{+}_{2},\nonumber\\
{\bf p}_{2\perp}&=&(1-x){\bf P}_{2\perp} +
{\bf k'}_{\perp},\hspace{.5cm}
{\bf p'}_{\bar{q}\perp}= x{\bf P}_{2\perp} - {\bf k'}_{\perp},
\end{eqnarray}
which requires that $p^{+}_{\bar{q}}=p'^{+}_{\bar{q}}$
and ${\bf p}_{\bar{q}\perp}={\bf p'}_{\bar{q}\perp}$. 
Our analysis for $K_{l3}$ decays will be carried out using this $q^{+}=0$ 
frame where the decaying hadron(Kaon) is at rest.
Using the matrix element of the ``$+$"-component of the current, $J^{+}$,
given by Eq.(2), we obtain the form factor $f_{+}(q_{\perp}^{2})$ as follows
\begin{eqnarray}
f_{+}(q_{\perp}^{2}) &=& \int^{1}_{0}dx\int d^{2}{\bf k}_{\perp}
\phi_{2}(x,{\bf k'}_{\perp})\phi_{1}(x,{\bf k}_{\perp})
\frac{{\cal A}_{1}{\cal A}_{2}+{\bf k}_{\perp}\cdot{\bf k'}_{\perp}}
{ \sqrt{ {\cal A}_{1}^{2}+ k^{2}_{\perp}}\sqrt{ {\cal A}_{2}^{2}+
k^{'2}_{\perp}} },
\end{eqnarray}
where $q_{\perp}^{2}=-q^{2}$, ${\cal A}_{i}=m_{i}x + m_{\bar{q}}(1-x)$ 
and ${\bf k'}_{\perp}={\bf k}_{\perp}-x{\bf q}_{\perp}$. As we discussed
in the introduction, we need the ``$\perp$"-component of the current, 
$J_{\perp}$, to obtain the form factor $f_{-}(q_{\perp}^{2})$ in Eq.(2), 
viz.,   
\begin{eqnarray}
<P_{2}|\bar{q}_{2}({\bf q}_{\perp}\cdot\vec{\gamma}_{\perp})q_{1}|P_{1}>
&=& q^{2}_{\perp}\biggl[f_{-}(q_{\perp}^{2}) -f_{+}(q_{\perp}^{2})\biggr], 
\end{eqnarray}
after multiplying ${\bf q}_{\perp}$ on both sides of Eq.(2).
The l.h.s. of Eq.(9) is given by
\begin{eqnarray}
<P_{2}|\bar{q}_{2}({\bf q}_{\perp}\cdot\vec{\gamma}_{\perp})q_{1}|P_{1}>
&=& -\int dxd^{2}{\bf k}_{\perp}
\frac{x\phi_{2}(x,{\bf k'}_{\perp})\phi_{1}(x,{\bf k}_{\perp})}
{ 2\sqrt{ {\cal A}_{1}^{2}+ k^{2}_{\perp}}\sqrt{ {\cal A}_{2}^{2}+
k_{\perp}^{'2} } }\nonumber\\
&\times&{\rm Tr} \biggl[\gamma_{5}({\not\! p}_{2}+m_{2})
({\bf q}_{\perp}\cdot\vec{\gamma}_{\perp})({\not\! p}_{1}+m_{1})\gamma_{5}
({\not\! p}_{\bar{q}}-m_{\bar{q}})\biggr].
\end{eqnarray}
Using the quark momentum variables given in Eq.(7), we obtain
the trace term in Eq.(10) as follows
\begin{eqnarray}
&& {\rm Tr} \biggl[\gamma_{5}({\not\! p}_{2}+m_{2})
({\bf q}_{\perp}\cdot\vec{\gamma}_{\perp})({\not\! p}_{1}+m_{1})\gamma_{5}
({\not\! p}_{\bar{q}}-m_{\bar{q}})\biggr]\nonumber\\
&=& -2\biggl\{ \frac{({\cal A}^{2}_{1}+ k_{\perp}^{2})}{x(1-x)}
({\bf k}_{\perp}-{\bf q}_{\perp})\cdot{\bf q}_{\perp} +
\frac{({\cal A}^{2}_{2}+ k_{\perp}^{'2})}{x(1-x)}
{\bf k}_{\perp}\cdot{\bf q}_{\perp}
+ \biggl[ (m_{1}-m_{2})^{2}
+ q_{\perp}^{2}\biggr]{\bf k}_{\perp}\cdot{\bf q}_{\perp}\biggr\}.
\end{eqnarray}
The more detailed derivation of Eqs.(8) and (10) are presented in 
appendix A. Since both sides of Eq.(9) vanish as $q^{2}\to 0$, one has to 
be cautious for the numerical computation of $f_{-}$ at $q^{2}=0$. Thus,
for the numerical computation at $q^{2}=0$, we need to find an analytic
formula for $f_{0}$. In order to obtain the analytic formula for the form 
factor $f_{-}(0)$, we make a low $q^{2}_{\perp}$ expansion to extract the 
overall $q^{2}_{\perp}$ from Eq.(10). 
Then, the form factor $f_{-}(0)$ is obtained as follows
\begin{eqnarray}
f_{-}(0)&=& f_{+}(0) + \int^{1}_{0}dx\int d^{2}{\bf k}_{\perp}
\frac{x\phi_{2}(x,{\bf k}_{\perp})\phi_{1}(x,{\bf k}_{\perp})}
{ \sqrt{ {\cal A}_{1}^{2}+ k^{2}_{\perp}}\sqrt{ {\cal A}_{2}^{2}+
k_{\perp}^{2} } }\nonumber\\
&\times&\biggl\{ \biggl[C_{T1}(C_{J1}-C_{J2}+C_{M}+C_{R}) + C_{T2}\biggr ]
k^{2}_{\perp}\cos^{2}\phi + C_{T3}\biggr\},
\end{eqnarray}
where the angle $\phi$ is defined by ${\bf k}_{\perp}\cdot{\bf q}_{\perp}
=|{\bf k}_{\perp}||{\bf q}_{\perp}|\cos\phi$ and the terms of $C's$ are 
given by 
\begin{eqnarray}
C_{Ji}&=&
\frac{2\beta^{2}_{i}}{(1-x)(\beta^{2}_{1}+\beta^{2}_{2})M^{2}_{i0}}
\biggl[\frac{1}{1 -[(m^{2}_{i}-m^{2}_{\bar{q}})/M^{2}_{i0}]^{2}}  
-\frac{3}{4}\biggr], \nonumber\\
C_{M}&=&\frac{1}{(1-x)(\beta^{2}_{1}+\beta^{2}_{2})}
\biggl[\frac{\beta^{2}_{2}}{M^{2}_{20}-(m_{2}-m_{\bar{q}})^{2}}
- \frac{\beta^{2}_{1}}{M^{2}_{10}-(m_{1}-m_{\bar{q}})^{2}}\biggr],\nonumber\\
C_{R}&=&\frac{-1}{4(1-x)(\beta^{2}_{1}+\beta^{2}_{2})}
\biggl[ \biggl(\frac{m^{2}_{2}-m^{2}_{\bar{q}}}{M^{2}_{20}}\biggr)^{2}
- \biggl(\frac{m^{2}_{1}-m^{2}_{\bar{q}}}{M^{2}_{10}}\biggr)^{2}\biggr],
\nonumber\\
C_{T1}&=& \frac{1}{x(1-x)}({\cal A}_{1}^{2}+ {\cal A}_{2}^{2}+
2k^{2}_{\perp}) + (m_{1}-m_{2})^{2},\nonumber\\
C_{T2}&=& \frac{2(\beta^{2}_{1}-\beta^{2}_{2})}
{(1-x)(\beta^{2}_{1}+\beta^{2}_{2})},\hspace{.1cm} 
C_{T3}= \frac{x\beta^{2}_{1}}{\beta^{2}_{1}+\beta^{2}_{2}}C_{T1}
-\frac{{\cal A}^{2}_{1}+k^{2}_{\perp}}{x(1-x)},
\end{eqnarray}
with 
\begin{eqnarray}
M^{2}_{i0} &=& \frac{k_{\perp}^{2}+m^{2}_{i}}{1-x}
+ \frac{k_{\perp}^{2}+m^{2}_{\bar{q}}}{x}.
\end{eqnarray}
The form factors $f_{+}$ and $f_{-}$ can be analytically continued to  
the time-like $q^{2}>0$ 
region\footnote{ We note that our numerical results of $f_{+}$ obtained
by the method of replacing $q_{\perp}$ by $iq_{\perp}$ in Eq.(8)
for any $P\to P$($P$=Pseudoscalar) semileptonic decays are identical to  
those obtained from dispersion formulation in Ref.\cite{Melikhov}.} 
by replacing $q_{\perp}$ by $iq_{\perp}$ in
Eqs.(8) and (9). Since $f_{-}(0)$ in Eq.(12) is exactly 
zero in the $SU_{f}(3)$ symmetry\cite{Yuri}, $i.e.$, $m_{u(d)}=m_{s}$ and 
$\beta_{u\bar{d}}=\beta_{u\bar{s}}=\beta_{s\bar{s}}$, one can get  
$f_{+}(q_{\perp}^{2})=F_{\pi}(q_{\perp}^{2})$ for the $\pi^{+}\to\pi^{0}$ 
weak decay($\pi_{e3}$), where $F_{\pi}(q_{\perp}^{2})$ 
is the electromagnetic form factor of pion, and $f_{-}(q^{2})=0$ because of 
the isospin symmetry. For comparison, we briefly discuss in appendix B 
the form factors in $q^{+}\neq 0$ frame.
\section{Numerical results}
As we discussed in the introduction, we used the same quark model 
parameters $(m_{u(d)},m_{s},\beta_{u\bar{d}},\beta_{u\bar{s}})$ 
as in Ref.\cite{CJ1} to predict various observables for $K_{l3}$ decays. 
These parameters are summarized in Table I. The Sets 1 and 2 in
this Table represent the model parameters obtained by  
the harmonic oscillator and linear confinement potentials, 
respectively, from Ref.\cite{CJ1}.  

Our predictions of the parameters for $K_{l3}$ decays in $q^{+}=0$ 
frame, i.e., $f_{+}(0)$, $\lambda_{+}$, $\lambda_{0}$, 
$<r^{2}>_{K\pi}=6f'_{+}(0)/f_{+}(0)= 6\lambda_{+}/M^{2}_{\pi^{+}}$, and 
$\xi_{A}=f_{-}/f_{+}|_{q^{2}=m^{2}_{l}}$,
are summarized in Table II. We do not distinguish $K_{e3}$ 
from $K_{\mu3}$ in the calculation of the above parameters 
since the slopes of $f_{\pm}$ are
almost constant in the range of $m^{2}_{e}\leq q^{2}\leq m^{2}_{\mu}$.   
However, the decay rates should be different due to the phase space 
factors given by Eq.(4) and our numerical results for $\Gamma(K_{e3})$
and $\Gamma(K_{\mu3})$ in $q^{+}=0$ frame are also presented in Table II. 
Our results for the form factor $f_{+}$ at zero momentum 
transfer, $f_{+}(0)=0.961[0.962]$ for set 1[set 2], are consistent with the 
Ademollo-Gatto theorem\cite{Gatto} and also in an excellent agreement
with the result of chiral perturbation theory\cite{Roos},
$f_{+}(0)=0.961\pm 0.008$. Our results for other observables such as
$\lambda_{+}$, $\xi_{A}$, and $\Gamma(K_{l3})$  
are overall in a good agreement with the experimental data\cite{data}. 
We have also investigated the sensitivity of our results by  
varying quark masses. For instance, the results\footnote{ Even though we
show the results only for the set 1, we find the similar variations for 
the set 2, i.e., the posivive sign of $\lambda_{0}$ can be obtained when
$m_{s}/m_{u}\leq 1.8$ for both sets 1 and 2. In addition to the observables
in this work, our predictions for $f_{K}$, $r^{2}_{K^{+}}$, and
$r^{2}_{K^{0}}$ in \cite{CJ1} are changed to 108 MeV($1\%$ change),
0.385fm$^{2}$($0.3\%$ change), and $-0.077$fm$^{2}$(15$\%$),
respectively.} obtained by changing the strange quark mass from 
$m_{s}=0.48$ GeV to 0.43 GeV($10\%$ change) for the set 1 are included in 
Table II.
As one can see in Table II, our model predictions are quite stable for the 
variation of $m_{s}$ except $\lambda_{0}$, which changes its sign from 
$-0.007$ to $+0.0027$. The large variation of $\lambda_{0}$ 
is mainly due to the rather large sensitivity of $f_{-}(0)$($18\%$
change) to the variation of $m_{s}$.
Similar observation regarding on the large sensitivity for $\lambda_{0}$ 
compared to other observables has also been reported in Ref.\cite{Andrei} 
for the variation of quark masses.    
As discussed in Refs.\cite{Yuri} and \cite{Isgur}, $f_{-}(0)$ is sensitive 
to the nonperturbative enchancement of the SU(3) symmetry breaking mass
difference $m_{s}-m_{u(d)}$ since $f_{-}(0)$ depends on the ratio of $m_{s}$
and $m_{u(d)}$. 

Of special interest, we also observed that the non-valence contributions 
from $q^{+}\neq0$ frame are clearly visible for $\lambda_{+}$, 
$\lambda_{0}$ and $\xi_{A}$ even though it may not be quite significant 
for the decay rate $\Gamma(K_{l3})$.
Our predictions with only the valence contributions in $q^{+}\neq 0$ 
frame are $f_{+}(0)=0.961[0.962]$, $\lambda_{+}=0.081[0.083]$,  
$\lambda_{0}=-0.014[-0.017]$, $\xi_{A}=-1.12[-1.10]$,
$\Gamma(K_{e3})=(8.02[7.83]\pm 0.13)\times10^{6}s^{-1}$ and
$\Gamma(K_{\mu3})=(4.49[4.36]\pm 0.13)\times10^{6}s^{-1}$ for the
set 1[set 2].  Even though the form factor $f_{+}(0)$ in $q^{+}\neq 0$
frame is free from the non-valence contributions, its derivative at $q^{2}=0$, 
i.e., $\lambda_{+}$, receives the non-valence contributions. Moreover, 
the form factor $f_{-}(q^{2})$ in $q^{+}\neq 0$ frame is not immune to
the non-valence contributions even at $q^{2}=0$\cite{Zm}.
Unless one includes the non-valence contributions in the $q^{+}\neq0$ frame,
one cannot really obtain reliable predictions for the observables such as 
$\lambda_{+}, \lambda_{0}$ and $\xi_{A}$ for $K_{l3}$ decays.

In Fig. 2, we show the form factors $f_{+}$ obtained from both $q^{+}=0$
and $q^{+}\neq0$ frames for $0\leq q^{2}\leq (M_{K}-M_{\pi})^{2}$ region. 
As one can see in Fig.2, the form factors $f_{+}$ 
obtained from $q^{+}=0$ frame(solid lines) for both parameter sets 1 and 2  
appear to be linear functions of $q^{2}$ justifying Eq.(6) usually employed 
in the analysis of experimental data\cite{data}. Note, however, that 
the curves without the non-valence contributions in  
$q^{+}\neq 0$ frame(dotted lines) do not exhibit the same behavior. 
In Fig.3, we show $d\Gamma/dq^{2}$ spectra for $K_{e3}$(solid line) and
$K_{\mu3}$(dotted line) obtained from $q^{+}=0$ frame.  
While the term proportional to $f_{0}$ in Eq.(4) is negligible for 
$K_{e3}$ decay rate, its contribution for $K_{\mu3}$ decay rate is quite 
substantial(dot-dashed line). Also, we show in Fig.4 the form factors 
$f_{+}(q^{2})$(solid and dotted lines for the sets 1 and 2, respectively)
at spacelike momentum transfer region and compare with the theoretical 
prediction from Ref.\cite{Andrei}(dot-dashed line).  
The measurement of this observable in $q^{2}<0$ region is anticipated 
from TJNAF\cite{Andrei}. 

In addition, we calculated the electromagnetic form factors
$F_{\pi}(q^{2})$ and $F_{K}(q^{2})$ in the space-like region using
both $q^{+}=0$ and $q^{+}\neq 0$ frames to estimate the non-valence
contributions in $q^{+}\neq 0$ frame.  
As shown in Figs.5 and 6 for $F_{\pi}(q^{2})$ and $F_{K}(q^{2})$,
respectively, our predictions in $q^{+}=0$ frame are in a very good
agreement with the available data\cite{Amen1,Amen2} while the results 
for $q^{+}\neq0$ frame deviate from the data significantly. The deviations 
represent the non-valence contributions in $q^{+}\neq 0$ 
frame(see Fig.1(b)).  However, the deviations are clearly
reduced for $F_{K}(q^{2})$(see Fig.6) because of the large suppression
from the energy denominator shown in Fig.1 for the non-valence
contribution. The suppressions are much bigger for the heavier mesons
such as $D$ and $B$. Especially, for the $B$ meson case, the
non-valence contribution is almost negligible
up to $Q^{2}=-q^{2}\sim 10$ GeV$^{2}$.
\section{Summary and Conclusion}  
In this work, we investigated the weak decays of $K_{l3}$ using the  
light-front quark model. The form 
factors $f_{\pm}$ are obtained in $q^{+}=0$ frame and then analytically 
continued to the time-like region by changing $q_{\perp}$ to 
$iq_{\perp}$ in the form factors. The matrix element of the 
``$\perp$"-component of the current $J^{\mu}$ is used to obtain 
the form factor $f_{-}$, which is necessary for the complete analysis of 
$K_{l3}$ decays.
Using the non-zero lepton mass formula[Eq.(4)] for the decay rate of 
$K_{l3}$, we also distinguish $K_{\mu3}$ from $K_{e3}$ decay. Especially, 
for $K_{\mu3}$ decay, the contribution from $f_{0}$[or $f_{-}$] form 
factor is not negligible in the calculation of the decay rate.  
Our theoretical predictions for $K_{l3}$ weak decays are overall in a good 
agreement with the experimental data. 
We also confirmed that our analytic continuation method is 
equivalent to that of Ref.\cite{Melikhov} where the form factors are
obtained by the dispersion representations through the (gaussian) wave 
functions of the initial and final mesons. 
In all of these analyses, it was crucial to include the non-valence 
contributions in $q^{+}\neq 0$ frame. As we have estimated these 
contributions in various observables, their magnitudes are not at all
negligible in the light-to-light electroweak form factors.
In fact, the non-valence contributions were very large for the most 
of observables such as $\lambda_{+}$, $\lambda_{0}$, $\xi_{A}$, 
$F_{\pi}(Q^{2})$ and $F_{K}(Q^{2})$.

Finally, we have also estimated the zero-mode contribution by calculating 
the $``-"$-component of the current. Our observation in an exactly 
solvable scalar field theory was presented in Ref.\cite{Zm}.
Using the light-front bad current $J^{-}$ in $q^{+}=0$ frame, 
we obtained $f_{-}(0)=12.6[18.6]$ for the set 1[set 2]. 
The huge ratio of $f_{-}(0)|_{J^{-}}/f_{-}(0)|_{J_{\perp}}\approx 
-36[-48]$ for the set 1[set 2]  
is consistent with our observation in Ref.\cite{Zm}. 
We also found that the zero-mode contribution is highly
suppressed as the quark mass increases.
The detailed analysis of heavy-to-heavy and heavy-to-light semileptonic 
decays is currently underway.
\acknowledgements 
This work was supported by the U.S. DOE under
contracts DE-FG02-96ER 40947. The North Carolina
Supercomputing Center and the National Energy Research Scientific
Computer Center are also acknowledged for the grant of supercomputer
time. We would like to acknowledge Andrei Afanasev for giving us 
their data for $K_{l3}$ so that we can compare our results with theirs.
%\newpage
%\noindent
%\appendix
%\setcounter{section}{0}
\setcounter{equation}{0}
\renewcommand{\theequation}{\mbox{A\arabic{equation}}}
\begin{center}
{\bf APPENDIX A: DERIVATION OF THE MATRIX ELEMENT OF THE WEAK 
VECTOR CURRENT $<P_{2}|\bar{q}_{2}\gamma^{\mu(=+,\perp)}q_{1}|P_{1}>$ IN 
$q^{+}=0$ FRAME}
\end{center}
In this appendix A, we show the derivation of the matrix element of the 
weak vector current $<P_{2}|\bar{q}_{2}\gamma^{\mu}q_{1}|P_{1}>$ given in
Eq.(2) for $\mu=+$ and $\perp$, respectively.  

In the light-front quark model, the matrix element of the weak vector
current can be calculated by the convolution of initial and final 
light-front wave function of a meson as follows 
\begin{eqnarray} 
<P_{2}|\bar{q}_{2}\gamma^{\mu}q_{1}|P_{1}>&=& 
\sum_{\lambda_{1},\lambda_{2},\bar{\lambda}}
\int dp^{+}_{\bar{q}}d^{2}{\bf k}_{\perp}  
\phi^{\dagger}_{2}(x,{\bf k'}_{\perp})\phi_{1}(x,{\bf k}_{\perp}) 
\nonumber\\ 
&\times&  
{\cal R}^{00\dagger}_{\lambda_{2}\bar{\lambda}}(x,{\bf k'}_{\perp})
\frac{\bar{u}(p_{2},\lambda_{2})}{\sqrt{p^{+}_{2}}}\gamma^{\mu}
\frac{u(p_{1},\lambda_{1})}{\sqrt{p^{+}_{1}}}  
{\cal R}^{00}_{\lambda_{1}\bar{\lambda}}(x,{\bf k}_{\perp}),      
\end{eqnarray}
where the spin-orbit wave function ${\cal R}^{JJ_{z}}(x,{\bf k}_{\perp})$ 
for pseudoscalar meson($J^{PC}=0^{-+}$) obtained from Melosh transformation 
is given by  
\begin{eqnarray} 
{\cal R}^{00}_{\lambda_{i}\bar{\lambda}}&=& 
\frac{1}{\sqrt{2}\sqrt{ M^{2}_{i0} -(m_{i}-m_{\bar{q}})^{2} }}       
\bar{u}(p_{i},\lambda_{i})\gamma^{5}v(p_{\bar{q}},\bar{\lambda}),  
\end{eqnarray}
and
\begin{eqnarray}
M^{2}_{i0} &=& \frac{k_{\perp}^{2}+m^{2}_{i}}{1-x}
+ \frac{k_{\perp}^{2}+m^{2}_{\bar{q}}}{x}.
\end{eqnarray}
Subsituting Eq.(A2) into Eq.(A1) and using the quark momentum variables
given in Eq.(7), one can easily obtain 
\begin{eqnarray}
<P_{2}|\bar{q}_{2}\gamma^{\mu}q_{1}|P_{1}>
&=& -\int dxd^{2}{\bf k}_{\perp}
\frac{\phi^{\dagger}_{2}(x,{\bf k'}_{\perp})\phi_{1}(x,{\bf k}_{\perp})}
{2(1-x)\prod^{2}_{i}\sqrt{ M^{2}_{i0} -(m_{i}-m_{\bar{q}})^{2} }}
\nonumber\\
&\times&{\rm Tr} \biggl[\gamma_{5}({\not\! p}_{2}+m_{2})
\gamma^{\mu}({\not\!p}_{1}+m_{1})\gamma_{5}
({\not\! p}_{\bar{q}}-m_{\bar{q}})\biggr], 
\end{eqnarray}
where we used the following completeness relations of the Dirac spinors 
\begin{eqnarray}
\sum_{\lambda_{1,2}}u(p,\lambda)\bar{u}(p,\lambda)&=&{\not\!p}+m,
\hspace{.5cm} 
\sum_{\lambda_{1,2}}v(p,\lambda)\bar{v}(p,\lambda)={\not\!p}- m.
\end{eqnarray}  
In the standard $q^{+}=0$ frame where the decaying hadron is at
rest, the trace terms in Eq.(A4)  
for the  ``+" and ``$\perp$"-components of the vector current
$J^{\mu}=\bar{q}_{2}\gamma^{\mu}q_{1}$, respectively, 
are obtaind as follows   
\begin{eqnarray}  
& &{\rm Tr} \biggl[\gamma_{5}({\not\! p}_{2}+m_{2})
\gamma^{\mu}({\not\!p}_{1}+m_{1})\gamma_{5}
({\not\! p}_{\bar{q}}-m_{\bar{q}})\biggr]\nonumber\\
& &= -4\biggl[ p^{\mu}_{1}(p_{2}\cdot p_{\bar{q}} + m_{2}m_{\bar{q}})
+ p^{\mu}_{2}(p_{1}\cdot p_{\bar{q}}+ m_{1}m_{\bar{q}}) 
+ p^{\mu}_{\bar{q}}(-p_{1}\cdot p_{2}+ m_{1}m_{2})\biggr]\nonumber\\
& & = -\frac{4P^{+}}{x}\biggl[ {\cal A}_{1}{\cal A}_{2} 
+ {\bf k}_{\perp}\cdot{\bf k'}_{\perp}\biggr], \hspace{.5cm}
\mbox{for}\,\mu= +  \\
& & = -2\biggl[ \frac{({\cal A}^{2}_{1}+ k_{\perp}^{2})}{x(1-x)}
({\bf k}_{\perp}-{\bf q}_{\perp}) + \frac{({\cal A}^{2}_{2}
+ k_{\perp}^{'2})}{x(1-x)}{\bf k}_{\perp} + [ (m_{1}-m_{2})^{2} 
+ q_{\perp}^{2}]{\bf k}_{\perp}\biggr], 
\hspace{.5cm} \mbox{for}\,\mu=\perp  
\end{eqnarray}  
where ${\cal A}_{i}=m_{i}x + m_{\bar{q}}(1-x)$
and ${\bf k'}_{\perp}={\bf k}_{\perp}-x{\bf q}_{\perp}$. Our convention
of the scalar product, $p_{1}\cdot p_{2}=(p^{+}_{1}p^{-}_{2} 
+ p^{-}_{1}p^{+}_{2})/2 - {\bf p}_{1\perp}\cdot{\bf p}_{2\perp}$ were
used to derive Eqs.(A6) and (A7) from the second line of the above
eqution. Substituing Eqs.(A6) and (A7) into Eq.(A4), we now obtain 
the matrix element of the weak vector current 
$<P_{2}|\bar{q}_{2}\gamma^{\mu}q_{1}|P_{1}>$ for $\mu=+$(see Eq.(8))
and $\perp$(see Eq.(10)) in $q^{+}=0$ frame, respectively.  
\setcounter{equation}{0}
\renewcommand{\theequation}{\mbox{B\arabic{equation}}}
\begin{center}
{\bf APPENDIX B: VALENCE CONTRIBUTIONS IN $q^{+}\neq 0$ FRAME}
\end{center}
For the purely longitudinal momentum transfer, $i.e.$, ${\bf q}_{\perp}=0$
and  $q^{2}=q^{+}q^{-}$, the relevant quark momentum variables are
\begin{eqnarray}
p^{+}_{1}&=&(1-x)P^{+}_{1},\hspace{.5cm}
p^{+}_{\bar{q}}=xP^{+}_{1},\nonumber\\
{\bf p}_{1\perp}&=&(1-x){\bf P}_{1\perp} + {\bf k}_{\perp},
\hspace{.5cm}
{\bf p}_{\bar{q}\perp}= x{\bf P}_{1\perp} - {\bf k}_{\perp},
\nonumber\\
p^{+}_{2}&=&(1-x')P^{+}_{2},\hspace{.5cm}
p'^{+}_{\bar{q}}=x'P^{+}_{2},\nonumber\\
{\bf p}_{2\perp}&=&(1-x'){\bf P}_{2\perp} +
{\bf k'}_{\perp},\hspace{.5cm}
{\bf p'}_{\bar{q}\perp}= x'{\bf P}_{2\perp} - {\bf k'}_{\perp},
\end{eqnarray}
where $x(x'=x/r)$ is the momentum fraction carried by the spectator
$\bar{q}$ in the initial(final) state. The fraction $r$ is given in
terms of $q^{2}$ as follows\cite{Demchuk,Grach,Simula,Cheng}
\begin{eqnarray}
r_{\pm}&=&\frac{M_{2}}{M_{1}}\biggl[\biggl(
\frac{ M^{2}_{1}+M^{2}_{2}-q^{2}}{2M_{1}M_{2}}\biggr)\pm
\sqrt{\biggl(\frac{ M^{2}_{1}+M^{2}_{2}-q^{2}}{2M_{1}M_{2}}\biggr)^{2}-1}
\biggr],
\end{eqnarray}
where the $+(-)$ signs in Eq.(B2) correspond to the daughter meson
recoiling in the positive(negative) $z$-direction relative to
the parent meson. In this $q^{+}\neq 0$ frame, one
obtains\cite{Demchuk,Grach,Simula,Cheng}
\begin{eqnarray}
f_{\pm}(q^{2})&=&\pm \frac{(1\mp r_{-})H(r_{+}) - (1\mp r_{+})H(r_{-})}
{r_{+}-r_{-}},
\end{eqnarray}
where
\begin{eqnarray}
H(r) &=& \int^{r}_{0}dx\int d^{2}{\bf k}_{\perp}
\phi_{2}(x',{\bf k}_{\perp})\phi_{1}(x,{\bf k}_{\perp})
\frac{{\cal A}_{1}{\cal A'}_{2}+ k^{2}_{\perp}}
{\sqrt{{\cal A}_{1}^{2}+ k^{2}_{\perp}}
\sqrt{{\cal A'}_{2}^{2}+ k_{\perp}^{2}}},
\end{eqnarray}
and ${\cal A'}_{i}= m_{i}x' + m_{\bar{q}}(1-x')$.
\newpage 
\begin{table}
\caption{Quark masses $m_{q}$[GeV] and gaussian parameters 
$\beta$[GeV] used in our analysis. $q$=$u$ and $d$.}
\begin{tabular}{cccccc}
& $m_{q}$ & $m_{s}$ & $\beta_{q\bar{q}}$ &
$\beta_{s\bar{s}}$& $\beta_{q\bar{s}}$\\
\tableline
Set 1& 0.25 & 0.48 & 0.3194 & 0.3681 & 0.3419 \\
Set 2& 0.22 & 0.45 & 0.3659 & 0.4128 & 0.3886 \\  
\end{tabular}
\end{table}
\begin{table}
\caption{
Model predictions for the parameters of $K_{l3}$ decay form factors 
obtained from $q^{+}=0$ frame. The charge radius $r_{\pi K}$ is obtained
by $<r^{2}>_{\pi K}=6f'_{+}(q^{2}=0)/f_{+}(0)$. As a sensitivity check,
we include the results in square brackets by changing $m_{s}=0.48$ to 
0.43 GeV for the parameter set 1. The CKM matrix used in the calculation of
the decay width(in unit of $10^{6}s^{-1}$) is 
$|V_{us}|=0.2205\pm0.0018$[9].}
\begin{tabular}{ccccr}
Observables & Set 1[$m_{s}=0.48\to0.43$]& Set2 & Other models 
& Experiment\\
\tableline
$f_{+}(0)$ & 0.961[0.974] & 0.962 & $0.961\pm0.008^{a}$, 
$0.952^{e}$,$0.98^{f}$,$0.93^{g}$ & \\       
$\lambda_{+}$& 0.025[0.029] & 0.026 
& 0.031$^{b}$,0.033$^{c}$,0.025$^{d}$
& $0.0286\pm0.0022[K^{+}_{e3}]$\\ 
& & &0.028$^{e}$,0.018$^{f}$,0.019$^{g}$ & \\  
& & & & $0.0300\pm0.0016[K^{0}_{e3}]$\\ 
$\lambda_{0}$& $-0.007[+0.0027]$ & $-0.009$ & $0.017\pm0.004^{b}$ 
0.013$^{c}$,0.0$^{d}$ & $0.004\pm0.007[K^{+}_{\mu3}]$\\
& & &0.0026$^{e}$,$-0.0024^{f}$,$-0.005^{g}$ & 
$0.025\pm0.006[K^{0}_{\mu3}]$\\  
$\xi_{A}$& $-0.38[-0.31]$& $-0.41$ & $-0.164\pm0.047^{b}$  
$-0.24^{c}$,$-0.28^{d}$ & $-0.35\pm0.15[K^{+}_{\mu3}]$\\
& & &$-0.28^{e}$,$-0.25^{f}$,$-0.28^{g}$ & $-0.11\pm0.09[K^{0}_{\mu3}]$\\ 
$<r>_{\pi K}$(fm) & 0.55[0.59] & 0.56 & 
0.61$^{b}$,0.57$^{e}$, 0.47$^{f}$,0.48$^{g}$ & \\ 
$\Gamma(K^{0}_{e3})$ & $7.30\pm 0.12[7.60\pm 0.12]$
& $7.36\pm 0.12$ & &$7.7\pm 0.5[K^{0}_{e3}]$\\
$\Gamma(K^{0}_{\mu3})$& $4.57\pm 0.07[4.84\pm 0.08]$& $4.56\pm 0.07$ & 
&$5.25\pm 0.07[K^{0}_{\mu3}]$\\
\end{tabular}
\end{table}
$^{a}$ Ref.\cite{Roos},$^{b}$ Ref.\cite{Gasser},$^{c}$ Ref.\cite{CL}, 
$^{d}$ Ref.\cite{VMD}, $^{e}$ Ref.\cite{Andrei}. 
$^{f}$ Ref.\cite{Yuri},$^{g}$ Ref.\cite{ISGW2}. 

\newpage 
\figure{\hspace{.2in} Fig.2. The form factors $f_{+}(q^{2})$ for the 
$K\to\pi$ transition in time-like momentum transfer $q^{2}>0$. The solid 
and dotted lines are the results from the $q^{+}=0$ and $q^{+}\neq 0$
frames for the parameter sets 1 and 2 given in Table I, respectively.
The differences of the results between the two frames are the measure of 
the non-valence contributions from $q^{+}\neq0$ frame.}
\figure{\hspace{.2in} Fig.3. The decay rates $d\Gamma/dq^{2}$ of
$K_{e3}$(solid line) and $K_{\mu3}$(dotted line) for the parameter 
set 1 in $q^{+}=0$ frame. The dot-dashed line
is the contribution from the term proportion to $f_{0}$ in Eq.(4)
for $K_{\mu 3}$ decay. The results for the set 2 are not much different 
from those for the set 1.}
\figure{\hspace{.2in} Fig.4. The form factors $f_{+}(q^{2})$ for the
$K\to\pi$ transition in spacelike momentum transfer $-q^{2}<0$. The 
solid and dotted lines are the results from the sets 1 and 2,
respectively. The dot-dashed line is the result from Ref.\cite{Andrei}.} 
\figure{\hspace{.2in} Fig.5. The EM form factor of pion for low   
$Q^{2}=-q^{2}$ compared with data\cite{Amen1}. The solid
and dotted lines are the results from the $q^{+}=0$ and $q^{+}\neq 0$
frames for the parameter sets 1 and 2, respectively.}
\figure{\hspace{.2in} Fig.6. The EM form factor of kaon 
compared with data\cite{Amen2}. The same line code as in Fig.5 is used.}
\end{document}